\documentclass[fleqn,twoside]{article}

\usepackage{espcrc2}
\usepackage{graphicx}
\usepackage{amssymb}
\usepackage[figuresright]{rotating}

\newcommand{\AmS}{{\protect\the\textfont2
  A\kern-.1667em\lower.5ex\hbox{M}\kern-.125emS}}

\title{Production of high-energy $\mu$ neutrinos from young neutron stars}

\author{G. F. Burgio\address[INFN]{INFN Sezione di Catania, \\
        Via S. Sofia 64, I-95123 Catania, Italy}%
        \thanks{Talk given by G.F. Burgio}
         and
         B. Link\address{Department of Physics, Montana State
         University, \\
        Bozeman, Montana 59717, USA}}

\begin{document}

\begin{abstract}
Young, rapidly rotating neutron stars could accelerate protons to
energies of
 $\sim 1$ PeV close to the stellar surface, which scatter
with x-rays from the stellar surface through the $\Delta$
resonance and produce pions. The pions
 subsequently decay to produce muon
neutrinos. We find that the
 energy spectrum of muon neutrinos
consists of a sharp rise at $\sim
 50$ TeV, corresponding to the onset
of the resonance, above which
 the flux drops as $\epsilon_\nu^{-2}$
up to an upper-energy cut-off
 that is determined by either kinematics
or by the maximum energy to
 which protons are accelerated. We predict
event rates as high as
 10-50 km$^{-2}$ yr$^{-1}$ from relatively
young, close neutron
 stars. Such fluxes would be detectable by IceCube.
\end{abstract}

\maketitle

\section{Introduction}
A new window in high-energy astronomy is opening as existing
neutrino detectors are improved and new ones are developed. Since
neutrinos produced in astrophysical systems are unimpeded by
interstellar matter on their way to Earth, detections will provide a
new way to study the highest energy phenomena in the Universe.
Astrophysical neutrinos are expected to arise in many environments
 in
which neutrinos are produced by the decay of pions created
 through
hadronic interactions ($pp$) or photomeson production
 ($p\gamma$).
Neutrinos may be produced by cosmic accelerators, like
 those in
supernova remnants \cite{pro98}, active galactic nuclei \cite{lm00},
micro-quasars \cite{dist02} and gamma-ray bursts \cite{wb97,dai01}.
To detect these neutrinos, several projects are underway to develop
large-scale neutrino detectors under water or ice. AMANDA-II
\cite{ah04}, in the South Pole, and Baikal \cite{ay06} are the two
neutrino telescopes currently running, whereas ANTARES \cite{ca03}
and NESTOR \cite{tz03} are under construction in the Mediterranean
Sea. Those telescopes belong to a first generation with instrumented
volume smaller than 0.02 $\rm km^3$. The IceCube detector
\cite{ha06}, with a volume of about 1 $\rm km^3$, is under
construction on the same site of AMANDA. The NEMO project
\cite{mig06} is in its starting phase, and will be a cubic
kilometer size detector located at Capo Passero, Southern Italy.

As neutrino astronomy comes of age, it is important to get some idea
of what the sources might look like to aid in their detection.
Recently, we proposed \cite{lb05,lb06} that young ($t_{age}\lesssim
10^5 \rm yr$) and rapidly-rotating neutron stars could be  intense
neutrino sources. Here we summarize this work.

\section{The Model}

Neutron stars have enormous magnetic fields ($> 10^{12}$ G,
typically) and high rotation rates (tens of Hertz), causing them to
act as very powerful unipolar generators. Charges stripped off the
highly-conductive surface are accelerated somewhere above the
stellar surface. As the charges are accelerated along magnetic field
lines, they emit curvature radiation, which scatters with the
magnetic field and produces an $e^+e^-$ cascade. The cascade
produces a radio beam along the magnetic field lines that open to
infinity. If the beam angle and direction to Earth are favorable,
the star can be detected as a {\em radio pulsar}, with one radio
pulse per rotation period (in some cases, two pulses, when emission from
both magnetic poles is seen).

If the stellar magnetic moment has a component anti-parallel to the
spin axis (half of neutron stars), ions will be accelerated off the
surface (see Fig.1). If energies of $\sim$ 1 PeV per proton are
attained, pions will be produced through photomeson production as
the protons scatter with surface x-rays, producing a beam of $\mu$
neutrinos with energies above $\sim 50$ TeV. Detection of such
neutrinos would be a fascinating discovery in its own right, and
would provide an invaluable probe of the physical conditions that
prevail in the magnetosphere of a neutron star.

For photomeson production to occur, ions must be accelerated to very
high energies close to the stellar surface. How large might the
accelerating potential be? Goldreich and Julian developed the first
model of a quasi-static magnetosphere \cite{gj69}. By assuming a
dipolar configuration with magnetic axis parallel to the rotation
axis, they showed that the potential drop {\em across} the field
lines of a pulsar with angular velocity $\Omega=2\pi/p$ (where $p$
is the period)  from the magnetic pole to the last field line that
opens to infinity is of magnitude
\begin{equation}
\Delta\Phi=\frac{\Omega^2 B R^3}{2c^2}\simeq 7\times 10^{18} B_{12}
R_6^3 p_{ms}^{-2} \mbox{ Volts} \label{phimax}.
\end{equation}
Here $B=10^{12}B_{12}\rm G$ is the strength of the dipole component
of the field at the magnetic poles, $R=10^{6}R_{6}$ cm is the
stellar radius and $p_{\rm ms}$ is the spin period in milliseconds.
In equilibrium (not realized in a pulsar), a co-rotating
magnetosphere would exist in the regions above the star in which
magnetic field lines close; the charge density would be $\rho_q
\simeq eZn_0\simeq B/pc$ (cgs units), where $n_0$ is the equilibrium
number density of ions that would short out the component of the
electric field along the magnetic field. Deviation from corotation
will lead to charge-depleted gaps somewhere above the stellar
surface, through which charges will be accelerated to relativistic
energies \cite{rs75,as79}. Suppose that charge depletion occurs near
the stellar surface, with a characteristic density $f_d n_0$, where
$f_d<1$ is an unknown depletion factor. If the neutron star is
young, its surface will emit in soft x-rays, and the protons in
accelerated nuclei will scatter with this radiation field. If the
protons are sufficiently energetic, they will exceed the threshold
for photomeson production through the $\Delta^+$ resonance. The
$\Delta^+$ quickly decays to a $\pi^+$, and muon neutrinos are
produced through the following channel:
\begin{equation}
p\gamma\rightarrow \Delta^+\rightarrow n\pi^+ \rightarrow
n\nu_\mu\mu^+ \rightarrow n \nu_\mu e^+ \nu_e\bar{\nu}_\mu.
\end{equation}
The proton energy threshold $\epsilon_p$ for $\Delta^+$ production
is given by
\begin{equation}
\epsilon_p\epsilon_\gamma\ge 0.3\ {\rm GeV}^2 f_g, \quad\quad
f_g\equiv (1-\cos\theta)^{-1},
\end{equation}
where $\epsilon_\gamma$ is the photon energy and $\theta$ is the
incidence angle between the proton and the photon in the lab frame.
Young neutron stars typically have temperatures of $T_\infty \simeq
0.1 \rm \ keV$, and photon energies $\epsilon_\gamma=2.8
kT_\infty(1+z_g)\sim 0.4$ keV, where $z_g\simeq 0.4$ is the
gravitational red-shift and $T_\infty$ is the surface temperature
measured at infinity. The proton threshold energy for the $\Delta$
resonance is then $\epsilon_{p,{\rm th}}\simeq T_{\rm
0.1keV}^{-1}f_g$ PeV, where $T_{\rm 0.1keV}\equiv
(kT_\infty/0.1\mbox{ keV})$. Therefore, if the potential {\em along
field lines} is only $\sim 1$\% of the full potential $\Delta\Phi$
{\em across field lines} in the equilibrium magnetosphere, protons
can reach the $\Delta$ resonance.

Before continuing, we mention that
our assumption that PeV protons can be produced in the stellar
magnetosphere is a strong one; it is generally thought that pair
production near the stellar surface will quench the field and limit
the potential drop along the field to $\sim 1$ TeV
\cite{sturrock71,cr77}. These calculations are subject to numerous
uncertainties, and apply only in equilibrium, not the case in a
neutron star magnetosphere. Though potential drops as high as $\sim
1$ PeV might not be attainable, we consider the possibility
sufficiently interesting to explore further and to possibly confirm
or refute with future neutrino observations. Even null results would
provide a valuable probe of the poorly-understood neutron star
magnetosphere, confirming a limit that so far has only a theoretical
basis.

Assuming that a
potential of order $\Delta\Phi$ is available for acceleration {\em
along} field lines, a necessary condition for the $\Delta$ resonance
to be reached is
\begin{equation}
B_{12} p_{\rm ms}^{-2} T_{\rm 0.1keV} \ge 3\times 10^{-4}.
\label{threshold}
\end{equation}
Taking $T_{\rm 0.1keV}=1$, typical of pulsars younger than $\sim
10^5$ yr, there are 10 known pulsars within a distance of 8 kpc that
satisfy this condition \cite{psrcat}, about half of which should
have positively-charged magnetic poles; these are potentially
detectable sources of $\mu$ neutrinos. The best candidates are young
neutron stars, which are usually rapidly spinning and hot.
\begin{figure}[!t]
\includegraphics[width=8cm]{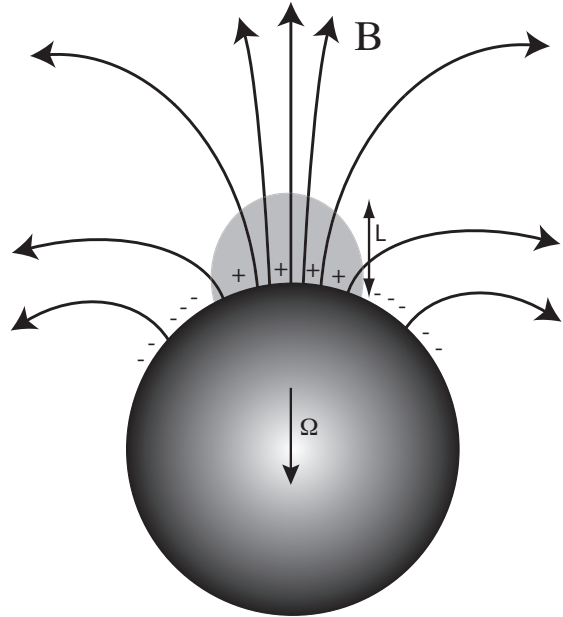}
\caption{For a neutron star with a component of its magnetic moment
antiparallel to its spin vector (these two vectors are shown here as
completely antiparallel for simplicity), the surface of the star
will be charged as shown. In the magnetic polar caps (one shown
here), positive charges will be stripped off the surface and
accelerated through a region of some dimension $L$ (shaded region).
We postulate that the charge depletion in this region is sufficient
that accelerating voltages of $\sim 1$ PeV per proton exist, at
least sporadically. The protons scatter with x-ray photons from the
stellar surface, producing pions through the $\Delta$
resonance.}\label{rotator}
\end{figure}
In the photomeson production process of eq.(2), the muon neutrinos
receive 5\% of the energy of the proton. Typical proton energies
required to reach resonance are $\sim 1$ PeV, so the expected $\mu$
neutrino energies will be $\sim 50$ TeV. The
conversion probability for a proton is small ($\lesssim 1$\%); only
a small number of protons are excited to the $\Delta$ resonance and
the pulsar beam is essentially unaffected. Moreover, since the
accelerated protons are far more energetic than the radiation field
with which they interact, any pions produced through the $\Delta$
resonance, and hence, any muon neutrinos, will be moving in nearly
the same direction as the proton was when it was converted. The
radio and neutrino beams will be approximately coincident, so that
some radio pulsars might also be detected as neutrino sources. We
see the radio beam for only a fraction $f_b$ of the pulse period,
the {\em duty cycle}. Typically, $f_b\simeq 0.1-0.3$ for younger
pulsars.  We take the duty cycle of the neutrino beam to be $f_b$
(but see below). In Ref. \cite{lb06}, we estimated the
phase-averaged neutrino flux at Earth resulting from the
acceleration of positive ions, at a distance $d$ from the source, to
be
\begin{equation}
\phi_\nu\simeq 2c f_b f_d(1-f_d) n_0 \left(\frac{R}{d}\right)^2 P_c,
\label{nuflux}
\end{equation}
where $f_dn_0$ is the space charge density in the depleted region,
where proton acceleration is taking place, and $P_c$ is the
conversion probability. For the purposes of calculating the
spectrum, we use the following differential neutrino flux:
\begin{equation}
\frac{d\phi_\nu}{d\epsilon_\nu}=2cf_bf_d(1-f_d)n_0\left(\frac{R}{d}\right)^2
\frac{dP_c}{d\epsilon_\nu}.
\label{diffflux}
\end{equation}
We are interested in obtaining upper limits on the flux and so
take $f_d=1/2$ and $Z=1$. The detailed calculation of the spectrum
$dP_c/d\epsilon_\nu$ has been reported in \cite{lb06}, and we
summarize these results next.

\begin{table*}[htb]
\caption{Estimated upper limits on the $\mu$ fluxes at Earth.
Numbers followed by question marks indicate guesses. Radio pulsar
spin parameters were taken from the catalogue of the Parkes
Radiopulsar Survey (www.atnf.csiro.au/research/pulsar/psrcat/).
Temperatures and limits on temperatures were taken from the
references indicated. The temperature upper limits on the Crab and
J0205+64 were used. The integrated conversion probability $P_c$ is
reported for the case of linear acceleration, and assuming $L=0.1$.}
\newcommand{\m}{\hphantom{$-$}}
\newcommand{\cc}[1]{\multicolumn{1}{c}{#1}}
\renewcommand{\tabcolsep}{0.9pc} % enlarge column spacing
\renewcommand{\arraystretch}{1.6} % enlarge line spacing
\begin{tabular}{@{}lllllllll}
\hline Source &{$d_{\rm kpc}$} & age & $p_{\rm ms}$ & $B_{12}$ &
$T_{\rm 0.1keV}$
& $f_b$ & $P_c$ & $dN/dAdt$ \\
 & & yr & & & &  & & km$^{-2}$ yr$^{-1}$   \\
\hline
Crab & 2 & $10^3$ & 33 & 3.8 & $\le 1.7$ \cite{wei04} & 0.14 & $\rm 1.6\times 10^{-3}$ & 45 \\
Vela & 0.29 & $10^{4.2}$ & 89 & 3.4 & 0.6 \cite{pav01}& 0.04 & $\rm 7.2\times 10^{-5}$ & 25 \\
J0205+64 & 3.2 & $10^{2.9}$ & 65 & 3.8 & $\le 0.9$ \cite{shm02}& 0.05 & $\rm 2.4\times 10^{-4}$ & 1 \\
B1509-58 & 4.4 & $10^{3.2}$ & 151 & 15 & 1? & 0.26 & $\rm 3.4\times 10^{-4}$ & 5 \\
B1706-44 & 1.8 & $10^{4.3}$ & 102 & 3.1 & 1? & 0.13 & $\rm 3.4\times 10^{-4}$ & 5 \\
B1823-13 & 4.1 & $10^{4.3}$ & 101 & 2.8 & 1? & 0.34 & $\rm 3.4\times 10^{-4}$ & 2 \\
Cass A & 3.5 & 300 & 10? & 1? & 4 \cite{pst04}& 0.1? &
$2.1\times 10^{-2}$ & 50 \\
SN 1987a & 50 & 17 & 1? & 1? & 4? & 0.1? & $2.1\times 10^{-2}$ &  3 \\
\hline
\end{tabular}
\end{table*}

\section{Neutrino Spectrum}
In calculating the neutrino spectrum, we have accounted for the fact that
the proton acceleration takes place over a finite distance above the
surface. We have considered an acceleration law of the form
\begin{equation}
\epsilon_p=\frac{\epsilon_0^2}{\epsilon_\gamma}\left(\frac{z}{L}\right)^\gamma
\end{equation}
where $\epsilon_0^2\equiv 0.3$ GeV$^2$, $z$ is the height above the
stellar surface and $L$ is the length scale over which the proton is
accelerated before reaching sufficient energy to undergo the
$\Delta$ resonance. We consider linear acceleration ($\gamma=1$),
corresponding to a constant electric field in the charge-depleted
zone, and quadratic acceleration ($\gamma=2$), corresponding to an
accelerating field that grows linearly with height. We have also
accounted for the finite width of the $\Delta$ resonance in the
cross section.

\begin{figure*}[!t]
\centering
 \includegraphics[width=8.cm, angle=270]{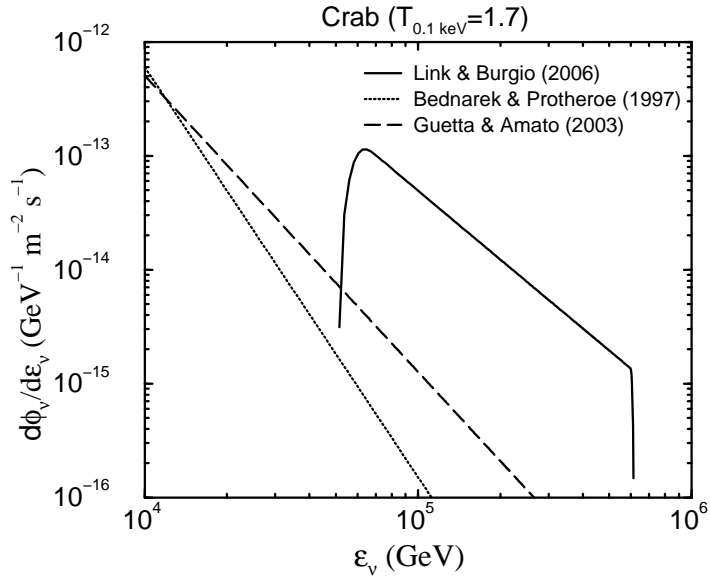}
  \caption{The predicted neutrino energy flux for the Crab
  pulsar for linear acceleration of the protons (solid line).
  Calculations from other models are also shown for comparison.}\label{spectrum}
\end{figure*}
\noindent

The spin parameters of 8 potential neutrino sources are given in
Table 1. In Fig. \ref{spectrum} we show the neutrino energy flux for
the Crab pulsar, assuming its surface temperature is equal to the
observational upper limit. The Crab pulsar is located in the
northern hemisphere and so is monitored by AMANDA-II. The spectrum
calculated in our model is represented by the solid line, and it has
been calculated by assuming linear acceleration. For comparison, we
also show two independent calculations obtained by
 Bednarek and Protheroe \cite{bp97} (dotted line), and Guetta and
 Amato \cite{ga03} (dashed line) in the case that neutrinos are
 emitted from the nebula. Those models predict a low muon event rate
 in comparison to our model, as shown in
Ref. \cite{bbm05}. In our model, the spectrum begins sharply at
\begin{equation}
\epsilon_\nu \simeq 70 T_{\rm keV}^{-1}\mbox{ GeV},
\end{equation}
corresponding to the onset of the resonance. At higher energies, the
spectrum drops approximately as $\epsilon_\nu^{-2}$, as the phase space for
conversion becomes restricted; higher energy neutrinos are produced by
protons that have been accelerated to greater heights, where the
photon density is lower and solid angle subtended by the star (as seen
by the proton) is smaller. At some maximum energy, the spectrum is
suddenly truncated by either kinematics (solid curve) or the
termination of the proton acceleration as limited by the magnitude of
the acceleration gap (not shown, since this cut-off cannot be
predicted).

\section{Estimated Count Rates and Discussion}

Large-area neutrino detectors use the Earth as a medium for
conversion of a muon neutrino to a muon, which then produces
\v{C}erenkov light in the detector. The conversion probability in
the Earth is $P_{\nu\mu \rightarrow \mu} \simeq 1.3 \times 10^{-6}
(\epsilon_{\nu}/\rm 1\,\, TeV)$, where $\epsilon_{\nu}$ is
the energy of the incident muon neutrino \cite{ghs95}. The muon event
rate is
\begin{equation}
\frac{dN}{dAdt} = \int d\epsilon_\nu \frac{d\phi_\nu}{d\epsilon_\nu}
P_{\nu_\mu \rightarrow \mu}. \label{rate}
\end{equation}
Estimated conversion probabilities for the Crab, Vela and 6 other
pulsars are given in Table 1 (column 8) for a characteristic
acceleration length $L=0.1$. The final column gives the estimated
event rates. Complete consideration of the kinematics gives event
rates are a factor of $\sim 10-30$ lower than estimated in our
previous paper \cite{lb05}. 

Neutrinos are produced at relatively high rates only if the protons
are accelerated through the resonance close to the star. In this
case, we obtain integrated count rates of several to $\sim 50$
km$^{-2}$ yr$^{-1}$.  Such count rates should be easily detected by
IceCube, and possibly by AMANDA-II or ANTARES with integration times
of about a decade (IceCube is planned to have replaced AMANDA-II by
then) for depletion factors of $f_d\simeq 1/2$. While the
characteristics of the spectrum presented here are robust, we caution
that the events rates we obtain are very rough upper limits, subject
to many uncertainties. For example, we have assumed that the neutrinos
are beamed into the same solid angle as the radio beam, which might
not be a correct assumption. The radio beam is thought to be produced
at about $10R$ \cite{cor78}. In our model, the pions are
produced much closer to the star. They then propagate to $\sim 1000R$
before decaying to neutrinos. At this distance from the star, the
field is unlikely to be dipolar, and it is difficult to say anything
definite about the distribution of pion velocities in this region. If
the neutrinos form a beam, the beam may be more or less collimated
than the radio beam. If it is more collimated, the neutrino event
rates could be higher than estimated here.

Results of 807 d of data from AMANDA-II are now available
\cite{gr05}. AMANDA-II has detected 10 events (over a background of
5.4) from the direction of the Crab pulsar, with energies higher
than 10 GeV. This result, though intriguing, is not statistically
significant; IceCube will be able to confirm or refute this result.
While it would be more exciting to see neutrinos from pulsars, the
accumulation of null results over the next decade would be
interesting as well; it would probably mean that photomeson
production is ineffective or non-existent in the neutron star
magnetosphere, thus providing a bound on the accelerating potential
that exists near the neutron star surface.

\label{lastpage}

\end{document}